\documentstyle[aps,epsf,graphicx]{revtex}

\newcommand{\beq}{\begin{equation}}
\newcommand{\eeq}{\end{equation}}        
\newcommand{\bqa}{\begin{eqnarray}}        
\newcommand{\eqa}{\end{eqnarray}}        

\newcommand{\ie}{{\frenchspacing\em i.\hspace{0.4mm}e.{}}}

\newcommand{\UKcol}{{\sf UKQCD collaboration}}
\newcommand{\Tr}{{\hbox{Tr}}}
\newcommand{\expt}[2]{\langle #2 \rangle_{#1}}

\begin{document}

\draft

\title{ 
\hfill\begin{minipage}{0pt}\scriptsize \begin{tabbing}
\hspace*{\fill} Liverpool LTH 427\\
\hspace*{\fill} Edinburgh 98/12\\
\end{tabbing} 
\end{minipage}\\[8pt]  
Tuning Actions and Observables in Lattice QCD
}

\author{\frenchspacing Alan C. Irving}
\address{Theoretical Physics Division,
         Department of Mathematical Sciences\\
         University of Liverpool,
         PO Box 147, Liverpool L69 3BX, UK}

\author{\frenchspacing James C. Sexton and Eamonn Cahill} 
\address{School of Mathematics,
         Trinity College,
         Dublin 2,
         Ireland}
\author{\frenchspacing Joyce Garden, B\'alint Jo\'o, Stephen M. Pickles
and Zbigniew Sroczynski}
\address{Department of Physics and Astronomy,
        James Clerk Maxwell Building,\\
        University of Edinburgh,
        Mayfield Road, Edinburgh EH9 3JZ,
        Scotland, UK}

\author{\UKcol{}}

\maketitle

\begin{abstract}

We propose a strategy for conducting lattice QCD simulations at fixed
volume but variable quark mass so as to investigate  the physical
effects of dynamical fermions. We present details of techniques which
enable this to be carried out effectively, namely the tuning in bare
parameter space and efficient stochastic estimation of the fermion
determinant. Preliminary results and tests of the method are presented.
We discuss further possible applications of these techniques.


\end{abstract}

\section{Introduction}

First results from full simulations of lattice QCD have  confirmed the
magnitude of the computational task ahead  and have shown only glimpses
of physics beyond the quenched approximation. A recent survey of results
can be found in~\cite{SGlat97}.

Preliminary results by the \UKcol{}~\cite{MTlat97,DFpaper}  using
an $O(a)$ improved action have shown a surprisingly strong dependence of
the effective lattice volume on the bare quark mass. This  complicates
chiral extrapolations of simulation measurements and obscures
comparisons with quenched calculations.  With this in mind, we
investigate how one might control the effective lattice volume by tuning
the bare action parameters while the effects of decreasing quark mass
are studied.

Before proceeding, we should clarify what we mean by \lq effective lattice
volume\rq{}. 
For definiteness,
consider the Wilson discretisation of QCD giving a lattice action
dependent on two bare parameters $\beta$ and $\kappa$ defined in the
usual way. 
In the quenched approximation, we have become used to thinking of
$\beta$ as uniquely controlling the lattice spacing $a$. 
At fixed $\beta$, one makes  lattice 
measurements of the rho mass or the Sommer scale $r_0$, 
defined by~\cite{R0}
\[
r_0^2{{dV}\over{dr}}\big|_{r_0}=1.65\, ,
\]
which 
is conceptually simpler for the present discussion
since, in this case, there is no need to consider
extrapolation in the (valence) quark mass.
One then matches the lattice value of $r_0$
to its physical value ($\approx 0.49$~fm as extracted 
from heavy quark spectroscopy~\cite{R0})  
to obtain the lattice spacing at that value of $\beta$.
This mapping $a_Q(\beta)$ between $\beta$ and physical lattice spacing is model
dependent in that it is unique to the quenched approximation.
The continuum limit is not accessible directly since the lattice
volume vanishes. One must remain at lattice spacing small enough
that discretisation errors are small but not so small that
finite volume effects are significant. 

There are two ways of extending these ideas in the presence of
dynamical fermions controlled by the bare mass parameter $\kappa$.
\begin{enumerate}
\item The conventional procedure is to construct a similar mapping between
$\beta$ and lattice spacing $a$ where the matching is made using the 
lattice value of $r_0$ {\em extrapolated} in $\kappa$ 
(sea quark mass) to the chiral limit. 
This yields a unique, but regularisation-dependent, mapping $a_{\chi}(\beta)$.
Comparisons with the continuum limit are made as in the quenched case.
\item Alternatively, one may consider matching the lattice 
value of $r_0$ at finite values of $\kappa$. Here, the 
picture is that the simulation is being done with sea quarks
of non-infinite mass. Each value of the bare quark mass (or $\kappa$)
then corresponds to a different approximation to 
continuum QCD with light dynamical quarks, in much the same way
as does quenched QCD (infinite $\kappa$).
Matching in this case results in a mapping $a(\beta,\kappa)$. 
Clearly, this is also regularisation-dependent.
\end {enumerate}
The term \lq effective lattice volume\rq{} refers to 
this second definition of lattice spacing.
Our proposal, then,
is to conduct simulations in which one attempts to
hold the effective lattice spacing fixed. 
In this way one is better able
to keep the physics constant, control lattice artifacts and 
finite volume effects while studying the effects of
light dynamical quarks.
In contrast, when adopting the first strategy, the significance
of lattice artifacts and finite volume effects changes as the 
chiral extrapolation is made. 

In order to carry out this programme one requires a practical way
of identifying curves in the $\beta,\kappa$ plane
of constant lattice spacing
\beq
a(\beta,\kappa)=\hbox{constant}\, .
\eeq

Consider $F$, the lattice measurement
of observable $f$ with dimension $d$, so that
\beq
F=fa^{-d}\, .
\eeq
Then, in the scaling region and to leading order, 
curves of constant $F$ yield
estimates of these curves of constant lattice spacing. 
In this way, one can track changes in $\beta$ required to 
compensate for changes in $\kappa$. As a simple example,
one can identify the $\beta$ shift involved in
comparisons of quenched and dynamical simulations.
In practice,
there will be residual dependence on the choice
of $F$. We would expect $r_0$ to be a \lq good\rq{} choice for
exposing sea quark dependence whereas $m_{\pi}$ would {\em not}, due to the 
strong dependence on valence quark mass and the
effects of chiral symmetry constraints.

In the rest of the paper, we show how the operator and action
matching technology introduced in~\cite{ACIJCS} can be used to
identify such curves. We demonstrate efficient algorithms for
achieving it and present some numerical tests.  
In section~\ref{s:curves} we summarise the relevant matching
formalism required and how it may be used.
In section~\ref{s:trln} we describe an
efficient algorithm for making stochastic estimates
of the fermion determinant~\cite{BFG,ACIlat97}. 
Results of numerical tests are presented in the next  section.
This is followed by a discussion of additional
applications of these techniques, including parallel
tempering simulations with dynamical fermions. Conclusions
and outlook are contained in the final section.

\section{Curves of constant physics}
\label{s:curves}
We first review the matching formalism introduced in~\cite{ACIJCS}.
Consider actions $S_1[U]$ and $S_2[U]$ describing two lattice gauge
theories with the same gauge configuration space $\{ U\}$
so that $(i=1,2)$
\begin{equation}
Z_i\equiv\int {\mathcal D}Ue^{-S_i[U]},\quad
<F>_i\equiv {1\over Z_i}\int {\mathcal D}U Fe^{-S_i}.
\label{eq:ZF}   
\end{equation}
For example, $S_1$ might be
the quenched Wilson action and $S_2$ the $O(a)$-improved action
for 2-flavour QCD~\cite{Clover}. In the present application, 
we will consider $S_1$ and $S_2$ to be the same improved fermion
action but at different points in the $\beta,\kappa$ plane.
Here, $F$ is some lattice observable.
Expectation values with respect to the two actions can be related
via a cumulant expansion whose leading behaviour implies\cite{ACIJCS}
\begin{eqnarray}
\label{eq:F2F1}
<F>_2 = <F>_1 +<\tilde{F}\tilde{\Delta}_{12}>_1 + \dots\\
\hbox{where}\quad\Delta_{12} \equiv S_1-S_2,\quad
\tilde{F}\equiv F - <F>\, \hbox{etc.}
\end{eqnarray}

In general, an action is a function of several parameters.
For example, the Wilson action depends on the bare parameters 
$\beta$ and $\kappa$.
In~\cite{ACIJCS} we considered matching action parameters 
in one of three distinct ways
\begin{itemize}
\item[M1:] match a given set of operators. i.e. require
$$<F_n>_1=<F_n>_2\, ;$$
\item[M2:] minimise the \lq distance\rq{} between
the actions, i.e. $\sigma^2(\tilde{\Delta}_{12})$;
\item[M3:] maximise the acceptance in an exact
algorithm for  $S_2$ constructed via accept/reject applied to
configurations generated with action $S_1$.
\end{itemize}
It was shown that, to lowest order, tuning prescriptions
M2 and M3 coincide. In fact, if the operators $F_n$ contribute
to the action with weights which are considered as tuning parameters,
then prescription M1 also coincides to lowest order.  
The prescriptions differ in a calculable way at next order.
Details are in~\cite{ACIJCS}.

In the present application we take
\beq
S_1 = S_{\rm eff}(\beta_0,\kappa_0)\quad\hbox{and}\quad
S_2 = S_{\rm eff}(\beta,\kappa)
\label{eq:Sparam}
\eeq
and seek to explore the bare parameter dependence of the lattice theory
using configurations generated at a series of
reference points $(\beta_0,\kappa_0)$ in parameter space.  

Here, $S_{\rm eff}$ is 
the effective action corresponding to lattice QCD with the 
fermions integrated out. \ie{}
\beq
S_{\rm eff}=-\beta W_{\Box}-T
\label{eq:Seff}
\eeq
where $W_{\Box}$ is the usual Wilson plaquette action
\beq
W_{\Box}\equiv\frac{1}{3}\sum_{\Box}{\rm Re}\Tr U_{\Box}
\eeq
and
\beq
T\equiv n_f\Tr\ln M[U]=\frac{n_f}{2}\Tr\ln (M^{\dagger}M)\, .
\eeq
The fermion matrix $M$ for the non-perturbative $O(a)$ improved theory 
is a function of both $\kappa$ and $\beta$. This is because the parameter
$c_{\rm sw}$~\cite{Clover} is a function of $\beta$~\cite{JansSomm}.
Since the improvement scheme fixes $c_{\rm sw}(\beta)$, one
must not treat $c_{\rm sw}$ as an independently tunable parameter.
However, as we shall see, the fact that the operator
$T(\beta,\kappa)$ is a function of $\beta$ as well as $\kappa$
introduces some practical complications. 

According to eqn.~\ref{eq:F2F1}, one requires measurements of
\beq
\Delta_{12}\equiv S_1 - S_2 = (\beta-\beta_0)W_{\Box}+
T(\beta,\kappa)-T(\beta_0,\kappa_0)\, 
\label{eq:kshift}
\eeq
in order to carry out parameter tuning. 
We discuss efficient algorithms for this in section~\ref{s:trln}.

For now, consider matching lattice observable $F$ at two
neighbouring points in the $\beta,\kappa$ plane:
\beq
(\beta_0,\kappa_0) \quad\hbox{and}\quad
(\beta,\kappa)\equiv(\beta_0+\delta\beta,\kappa_0+\delta\kappa)\, .
\eeq 
According to prescription M1 above and (\ref{eq:F2F1})
we require, to first order in small quantities,
\beq
<\tilde{F}\tilde{\Delta}_{12}>_1=0\, .
\label{eq:M1match}
\eeq
From this we can deduce that the constant $F$ curve is given by 
\beq
{{\delta\beta}\over{\delta\kappa}}=
-{{<\tilde{F}\delta\tilde{T}>_1}\over
{\delta\kappa}<\tilde{F}\tilde{W}_{\Box}>_1}
\label{eq:Fcurve}
\eeq
where
\beq
\delta\tilde{T}=\tilde{T}(\beta_0+\delta\beta,\kappa_0+\delta\kappa)-
	\tilde{T}(\beta_0,\kappa_0)\, .
\eeq
Equation (\ref{eq:Fcurve}) amounts to a  non-linear differential
equation since the right hand side involves $\delta\beta$ (via the
$c_{\rm sw}$ parameter). Linearising and taking the limit
yields for the constant $F$ curve,
\beq
{{d\beta}\over{d\kappa}}
=-{{
{{\partial<\tilde{F}>}\over{\partial\kappa}}
}\over{
{{\partial<\tilde{F}>}\over{\partial\beta}}
}}
=-{{<\tilde{F}{{\partial\tilde{T}}\over{\partial\kappa}}>}
\over{
<\tilde{F}(\tilde{W}_{\Box}+
{{\partial\tilde{T}}\over{\partial c_{\rm sw}}}\dot{c}_{\rm sw})>
}}\, .
\label{eq:dbdkM1}
\eeq  
The quantity 
\[
\dot{c}_{\rm sw}={{dc_{\rm sw}}\over{d\beta}}
\]
is well determined~\cite{JansSomm} and so the 
determination of constant $F$ curves reduces to measuring
correlations of the form
\beq
<\tilde{F}\tilde{W}_{\Box}>_1\quad{\rm and}\quad
<\tilde{F}\delta\tilde{T}>_1\, .
\label{eq:dfcor}
\eeq
The details of this will be described in section~\ref{s:tests}.

As pointed out in the Introduction, the details of these
curves will depend on the choice of $F$. For sensible choices
and reasonably 
physical values of the parameters, one would hope that
the corresponding curves of constant $a$, $a_F(\beta,\kappa)$ say,
would agree rather closely, locally at least.
For demonstration purposes, we will consider in section~\ref{s:tests}
several simple choices for $F$:
\begin{itemize}
\item $P$, the average plaquette, proportional to $W_{\Box}$.
\item $W_L$, various Wilson loops. 
\item $S_{\rm eff}$, the complete effective action itself.
\item Correlation matrices for measuring the static potential and $r_0$.
\item Hadron correlators. 
\end{itemize}
The first of these, $P$, may be readily measured with high accuracy
and so is excellent for testing the basic technology. 
However, it is not expected to shed much light on lattice spacing.
The last two are computationally more demanding but more relevant
to the project at hand, identifying curves of fixed physical volume.

One might expect that matching the full action ($F=S_{\rm eff}$)
would be more physically relevant than matching
the plaquette piece of the action. From (\ref{eq:dbdkM1})
we see that this curve is determined by correlations of the form
\beq
<\tilde{T}\delta\tilde{T}>_1
\label{eq:dScor}
\eeq
in addition to those of (\ref{eq:dfcor}). As we shall see in 
section~\ref{s:tests}, it is more difficult to obtain
unbiased estimators for these.

Now consider matching scheme M2 where \lq distances\rq{} in the
action space are minimised.
One can think of this as
defining \lq geodesics\rq{} 
in the $\beta,\kappa$ plane with respect to the metric
implied by (\ref{eq:ZF}) i.e.
\[
g_{\mu\nu}=<(\partial_{\mu}\tilde{S})(\partial_{\nu}\tilde{S})>\, .
\]
The corresponding affine connection would be
\[
\Gamma_{\mu\alpha\beta} =
<(\partial_{\mu}\tilde{S})(\partial_{\alpha\beta}\tilde{S})>\, .
\]
Simple minimisation yields, to first order, 
\beq
{{d\beta}\over{d\kappa}}=-
{{{<(\tilde{W}_{\Box}+{{\partial\tilde{T}}\over{\partial c_{\rm sw}}}
\dot{c}_{\rm sw}}){{\partial\tilde{T}}\over{\partial\kappa}}>}
\over{
<(\tilde{W}_{\Box}+{{\partial\tilde{T}}\over{\partial c_{\rm sw}}}
\dot{c}_{\rm sw})^2>
}}\, .
\label{eq:dbdkM2}
\eeq  

In section~\ref{s:applics} we show that these curves are
directly relevant to
simulations of full QCD using parallel tempering.
 Again, we note that these curves involve operator
correlations which are more complex to estimate.

Finally in this section, we observe that some of the above formalism
simplifies considerably in the case of the unimproved Wilson action 
($c_{\rm sw}=0$) since then, for example,
\beq
{{\partial \tilde{T}}\over{\partial\beta}}=
<\tilde{F}\tilde{W}_{\Box}>\, .
\label{eq:dFdb}
\eeq  

\section{Stochastic estimator of the fermion determinant}
\label{s:trln}
We require an unbiased estimator for $T=\hbox{TrLn}H$ where $H=M^\dagger
M$ is a hermitian positive-definite matrix. 
We will also require estimators for $T^2$, $\delta T$ and 
$T\delta T$. 
Bai, Fahey and Golub
\cite{BFG} have recently proposed estimators, with bounds,
for quantities of the form
\beq
u^\dagger g(H)v
\label{eq:ufv}
\eeq
where $g$ is some matrix function.
In our application $g$ is the logarithm and, for convenience of notation,
we set
\beq
L\equiv \hbox{Ln(H)}\, .
\eeq
Taking $u=v=\phi_i$, some normalised noise vector (e.g. $Z_2$ or Gaussian),
we can obtain a stochastic estimate of $T$
via
\beq
E_T={1\over{N_{\phi}}}\sum_{i=1}^{N_\phi}\phi_i^{\dagger}L\phi_i\, .
\label{eq:ET}   
\eeq
The corresponding variance of this estimator is
\beq
\sigma^2(E_T)={1\over{N_{\phi}}}\Tr(L^2)
\label{eq:varET}
\eeq
for complex Gaussian noise, and something less than this for $Z_2$ noise
($\pm 1$ on each of the complex component).
In the case of Gaussian noise, we also obtain rather directly
an efficient unbiased estimator for $T^2$,
\beq
E_{T^2}=(E_T)^2-{1\over{N_{\phi}}}E_Q
\label{eq:ET2}
\eeq
where $E_Q$ is an unbiased estimator for $Q=\Tr L^2$,
\beq
E_Q={1\over{N_{\phi}}}\sum_{i=1}^{N_\phi}\phi_i^{\dagger}L^2\phi_i\, .
\label{eq:EQ}
\eeq
In the case of $Z_2$ noise, the corresponding estimator 
is not readily accessible via the techniques described below, so we 
restrict the discussion to complex Gaussian noise.

In a companion paper~\cite{CSI}, we give fuller details of 
methods for evaluating the quantities $\phi_i^{\dagger}L\phi_i$,
$\phi_i^{\dagger}L^2\phi_i$ and so on. In practice, we make no use
of the bounds presented in~\cite{BFG}. Instead we use large enough
Lanczos systems so that the numerical convergence renders the bounds
irrelevant. 
The efficiency of the method results from
an elegant relationship between the nodes and weights required for
a $N$-point Gaussian quadrature and the eigenvalues and eigenvectors
(so-called Ritz pairs) of a Lanczos matrix of dimension $N$ \cite{BFG}.
In the companion paper~\cite{CSI}, we show that this relationship, 
and resulting accuracy remain good even when orthogonality is lost.
This is an important point for our application. If the
Lanczos system is large enough to avoid truncation errors,
one is well into the regime where orthogonality is lost in
standard numerical Lanczos methods. In the present paper we 
merely summarise the formulae required to 
obtain the present set of results. Preliminary results using 
these techniques were presented in~\cite{ACIlat97}.

The actual estimator which we use for $T\equiv \hbox{Tr} L$ is 
\beq
\hat{E}_T={1\over{N_{\phi}}}\sum_{i=1}^{N_\phi} I(\phi_i)
\label{eq:EhatT}   
\eeq
where
\beq
I(\phi_i) = \sum_{j=1}^N\omega_j^2\hbox{Ln}(\lambda_j)\, .
\label{eq:Iphi}   
\eeq
Here $\{\lambda_j^i\}$ $(j=1,2\dots N)$ are the eigenvalues of the 
$N$-dimensional tridiagonal Lanczos matrix formed using $\phi_i$ 
as a starting vector.
The weights $\{\omega_j^2\}$ are related to the corresponding 
eigenvectors~\cite{BFG}. In fact, $\omega_j$ is just the
first component of the $j$th eigenvalue of the tridiagonal
matrix.

The estimator $\hat{E}_Q$, for $Q\equiv\Tr L^2$, is obtained
from (\ref{eq:EhatT}) and (\ref{eq:Iphi}) using $\ln(\lambda_j^2)$
rather than $\ln(\lambda_j)$. 
For $T^2$ (see~(\ref{eq:ET2})), we define
\beq
\hat{E}_{T^2}=(\hat{E}_T)^2-{1\over{N_{\phi}}}\hat{E}_Q
\label{eq:EhT2}
\eeq

It is straightforward to show that, with the above definitions
\beq
 \expt{\phi}{\hat{E}_T}\approx \expt{\phi}{E_T} = T
\eeq
and
\beq
 \expt{\phi}{\hat{E}_{T^2}}\approx \expt{\phi}{E_{T^2}} = T^2  \, .
\eeq

The above Lanczos-based methods for evaluating
$\phi_i^{\dagger}L\phi_i$ are significantly more
efficient than Chebychev-based methods used previously
\cite{SW1,ACIJCS}. For a given level of accuracy in the 
present applications, they are
between 3 and 5 times more economical in the number
of matrix multiplications required.
Typically we achieve  six figure convergence of the quadrature with 70 
Lanczos steps on a matrix with a condition number ($\lambda_{\rm max}/
\lambda_{\rm min}$) of order $10^4$ or $10^5$.

Our goal was to achieve variance with respect to $\phi$ (see~(\ref{eq:varET}))
which was one order of magnitude less than that with respect to
the physical (gauge) distribution. We found that $N_{\phi}=80$ was
a suitably conservative number of noise vectors to use.

For estimating $\delta T$, we note that
\beq
E_{\delta T}=E_{T'}-E_T=
{1\over{N_{\phi}}}\sum_{i=1}^{N_\phi}\phi_i^{\dagger}(L'-L)\phi_i\, .
\label{eq:EdT}
\eeq
Thus we can achieve variance
\beq
\sigma^2(E_{\delta T})={1\over{N_{\phi}}}\Tr((L'-L)^2)
\label{eq:varEdT}
\eeq
if we use
\beq
\hat{E}_{\delta T}=\hat{E}_{T'}-\hat{E}_T
\label{eq:EhdT}
\eeq
provided we have employed the {\em same} set of
noise vectors. This is simple to arrange. 

In fact, one could use stochastic estimators of the 
form~(\ref{eq:ufv}) to estimate directly 
\beq
{{\partial T}\over{\partial\kappa}}=
{{n_f}\over{\kappa}}\Tr[1-M^{-1}]
={{n_f}\over{\kappa}}{\rm Re}\Tr[1-M(M^{\dagger}M)^{-1}]\, .
\label{eq:dTdk}
\eeq
An estimator for $\Tr[M(M^{\dagger}M)^{-1}]$ is obtained by setting
\beq
v=\phi\, ,\qquad u=M^{\dagger}\phi
\label{eq:uv}
\eeq
where $\phi$ is a suitable (e.g. Gaussian) noise vector.
Then
\beq
u^{\dagger}(M^{\dagger}M)^{-1}v=
v^{\dagger}M(M^{\dagger}M)^{-1}v
\label{eq:uvest}
\eeq 
is an unbiased estimator of  $\Tr[M(M^{\dagger}M)^{-1}]$ as
required. For this unsymmetric case ($u\neq v$), two Lanczos systems must
be used and a subtraction performed~\cite{BFG}. For the present analysis we 
have used the symmetric formalism as described above.
Further analysis and 
discussion of these and related stochastic estimators is presented 
in~\cite{CSI}.

In the next section we report results of some tests
of the matching procedures using the above algorithms.

\section{Numerical tests of matching}
\label{s:tests}

\subsection{Work estimates}
Having generated an ensemble of decorrelated configurations, one can
consider estimating the numerical derivatives required
for matching (see~(\ref{eq:dbdkM1})) in one of two ways: 
\begin{enumerate}
\item conduct two further simulations at neighbouring points in 
parameter space and make (uncorrelated)  measurements of $<F>$ on
each of these 
\item apply the stochastic trace log techniques of the previous section
to the existing ensemble.
\end{enumerate}

One can estimate the relative amount of work involved in these two
approaches. Suppose we seek to achieve an absolute error of $\epsilon$ on a
measurement of $\delta<F>$ by each method and that the variances of $<F>$
and of $<\delta\tilde{T}\tilde{F}>$ are $\sigma^2_F$ and
$\sigma^2_{\delta F}$ respectively. The ratio of work required for the
two approaches is then
\beq
{{W_2}\over{W_1}}=
{{n_T}\over{4}}\cdot
{{\sigma_{\delta F}^2}\over{\sigma_F^2}}\cdot
{{W_T}\over{W_{\rm HMC}}}
\label{eq:workrat1}
\eeq
where $W_T$ is the work done in a stochastic estimate of trace log 
on one configuration, $n_T$ is the number
of trace logs required (either 2 or 3) and 
$W_{\rm HMC}$ is the work done in generating
one decorrelated configuration by Hybrid Monte Carlo (HMC). 
In turn, we estimate
\beq
{{W_T}\over{W_{\rm HMC}}}=
{{N_{\phi}N_{\rm Lanc}}\over{N_{\rm step}N_{\rm solve}}}\cdot
{{A}\over{\tau_{\rm AC}}}
\label{eq:workrat2}
\eeq
where the various parameters are associated with HMC simulation and
the stochastic estimation of the determinant. 
These are defined in Table~\ref{tb:work} which also shows typical
values from the analysis presented below. With the numbers shown,
the ratio $W_T/W_{\rm HMC}$ is about
$1/80$. In Table~\ref{tb:vars} we show sample values of the
variances $\sigma_F^2$ and $\sigma^2_{\delta F}$ taken from the
present study. Putting all this together we estimate
\beq
{{W_2}\over{W_1}}\approx 0.045\, .
\label{eq:workrat3}
\eeq
Although the variance of 
$<\delta\tilde{T}\tilde{F}>$ 
can be quite large on large lattices,
it is proportional to $(\delta\kappa)^2$, or whatever the
relevant difference parameter is.
It is therefore possible, in principle, to
obtain acceptable accuracy for the relevant derivative
by method (2) with significantly less work.
The above example suggests this is  less than $5\%$ of that required
to obtain comparable accuracy by making additional simulations.
One can easily refine the above treatment to take account
of the work involved in equilibration. Of course, this makes method (2)
seem even more attractive.

\subsection{Fixed plaquette curve}
First, we present results from matching the average plaquette $<P>$.
This provides a check on some basic features of the procedures:
the first order truncation of the 
cumulant expansion~(\ref{eq:F2F1}) and the Lanczos-based noisy 
estimator algorithm~(\ref{eq:EhatT}).
The initial reference point in the $\beta,\kappa$ plane was taken as
\beq
(\beta_0,\kappa_0)=(5.2,0.1340)
\label{eq:b0k0}
\eeq
where a sequence of well-equilibrated configurations had been generated by
hybrid Monte Carlo
on a $8^324$ lattice
with the non-perturbatively improved action 
for 2 flavours~\cite{DFpaper}. 
At $\beta=5.2$, the
relevant improvement coefficient is $c_{\rm sw}=2.0171$~\cite{JansSomm}.
It is estimated from preliminary spectroscopy measurements on
larger lattices that $\kappa_{crit}$ is around $0.1365$ at this $\beta$
and $c_{\rm sw}$.
The values of $\kappa$ considered in this analysis are typical
of those being used in production simulations.
Plaquette measurements from all trajectories 
within this data sample showed 
autocorrelation times less than 20 trajectories.
Longer runs showed an autocorrelation time for the plaquette of roughly
double this. Bootstrap with binning was used to estimate the errors on
all quantities.

Using the algorithm and choices described in section~\ref{s:trln}, we
made stochastic estimates of $T\equiv\Tr\ln M^{\dagger}M$ on a sequence of
40 configurations separated by 20 trajectories. 
In order to make use of~(\ref{eq:Fcurve}) and~(\ref{eq:dbdkM1}) this must be
done using a minimum of $n_T=3$ parameter sets:
\beq
\{c_{\rm sw}(\beta_0),\kappa_0\},\quad
\{c_{\rm sw}(\beta_0),\kappa_0+\delta\kappa\},\quad
\{c_{\rm sw}(\beta_0+\delta\beta),\kappa_0\}\, .
\eeq
Recall that for the unimproved Wilson action, there is no need to
account for the additional $\beta$ dependence in $T$ which enters via 
$c_{\rm sw}(\beta)$.

Having selected a change in bare quark mass $\delta\kappa$  (for
example $\delta\kappa=-0.0005$),  we then use~(\ref{eq:dbdkM1}) with
$F\equiv P$ to estimate $d\beta/d\kappa$. 
This yields a first estimate of
the change $\delta\beta$ required to maintain a fixed value of $<P>$.
One can then use~(\ref{eq:Fcurve}) to verify this estimate of 
$\delta\beta$ by making further stochastic estimates of $T$ at
the final parameter set 
$\{c_{\rm sw}(\beta_0+\delta\beta),\kappa+\delta\kappa\}$.
In all cases studied, this last verification step has been well
satisfied and so one can in fact identify the matching curve
directly from the two partial derivatives as proposed in the previous 
section.
Results are shown in Table~\ref{tb:avP}.

We have then generated further dynamical fermion configurations
at the matched point
\beq
(\beta_0 +\delta\beta,\kappa_0+\delta\kappa)=(5.220,0.1335)
\label{eq:bk}
\eeq
and accumulated a similar ensemble of configurations for subsequent 
measurement. The corresponding value of $c_{\rm sw}$ is $1.9936$.     
The plaquette measurements were made using relatively high statistics
yielding the statistical errors shown. In order to make a
proper comparison, one should fold into the error on $<P>$
at $(5.220,1.9936,0.1335)$ that due to the uncertainty in $\beta$
($\pm 0.0010$). This would feed through to an additional uncertainty
in $<P>$ of $\pm 0.0004$. Thus the matching test is very well
satisfied for the plaquette.

The matching prediction done in reverse, back from
$\kappa=.1335$ to $.1340$, is also seen to be well satisfied.
From Table~\ref{tb:avP} we see that $(5.220,0.1335)$ is expected
to match with $(5.205(15),0.1340)$ in good agreement with
$(5.2,0.1340)$ from where the matching estimates were originally made. 
Also in Table~\ref{tb:avP}, 
we show the estimated $\beta$ shift 
corresponding to a further change of $-0.0005$ in $\kappa$. Steps of
this kind allow one to set up a grid of points from which which one can
then deduce the fixed plaquette curve in the $\beta,\kappa$ plane.

For the sake of completeness, we also show in Table~\ref{tb:avP} the
estimated shift required to match with quenched measurements of
$<P>$ (\ie{} at infinite $\kappa$). Independent gauge simulations
at $\beta=5.61$ show a good match of $<P>$ when one takes into account
the additional uncertainty in $<P>$ of around $\pm 0.009$ which would
feed through from the error of $\pm0.03$ on estimating $\beta$.
Of course, for such
large shifts the first order approximation may not be sufficiently
accurate. We have calculated the second order approximation to
$\delta\beta$~\cite{ACIJCS} for this quantity (+0.52(10)) but the statistical
error is such that one cannot reliably discriminate it from
the first order result (+0.41(3)) 
with the present level of statistics.

A further plaquette matching test is given in Table~\ref{tb:avP}.
In this example, 
the reference point for the \lq constant\rq{} plaquette
curve was $(5.2,0.1330)$. Again, direct simulation 
showed that the matching was accurate and self-consistent. 
The matched points on this curve ($<P>=0.5197(3)$), have been used to conduct
tests of parallel tempering as described in section~\ref{s:applics}.

\medskip
\subsection{Full action matching}
We have noted in section~\ref{s:curves} that some of the correlations required
to identify curves of constant action~(\ref{eq:dbdkM2}) are not directly
calculable by the techniques of section~\ref{s:trln}. Those
of the form
\beq
<T(c_{\rm sw},\kappa)T(c_{\rm sw}',\kappa')>
\eeq
require some care when setting up unbiased estimators. In particular,
\beq
E_{TT'}\equiv E_TE_{T'}
\label{eq:ETET1}
\eeq
 is {\em not} unbiased. For an unbiased estimator,
one requires something like that used for
$T^2$, \ie{} $E_{T^2}$. 
Unfortunately it is not so easy to evaluate 
the analogue of~(\ref{eq:EQ}) via the above Lanczos methods. However,
provided the variance of $E_T$ with respect to noise~(\ref{eq:varET})
is indeed small compared to that with respect to gauge fluctuations,
the estimator (\ref{eq:ETET1}) provides a useful approximation.
Using this approximation,
we have measured the shift $\delta\beta$ corresponding to
a shift $\delta\kappa$ at fixed action to compare with that
at fixed plaquette. For the first test shown in Table~\ref{tb:avP}, we find
$\delta\beta=0.0200(10)$ (statistical error only) 
consistent with the value $0.0199(10)$ found
for the fixed plaquette curve. In each case studied, we have found
such consistency. 

We conclude that the fixed action and fixed plaquette curves
are not significantly different at this order. 

\medskip
\subsection{Gauge invariant loops}
We are especially interested in matching those observables
 which are more sensitive to long range
physics. We have repeated the above plaquette matching analysis 
using a variety of Wilson loops.
The 16 loops used consist of 4 basic shapes realised in 4
different \lq magnifications\rq{} ($\times 1$, $\times 2$, $\times 3$ 
and $\times 4$). 
The 4 basic shapes used were those specified in terms of link steps by
the operations shown in Table~\ref{tb:loops} and rotations thereof.

\begin{figure}
\centerline{\includegraphics[width=\columnwidth]{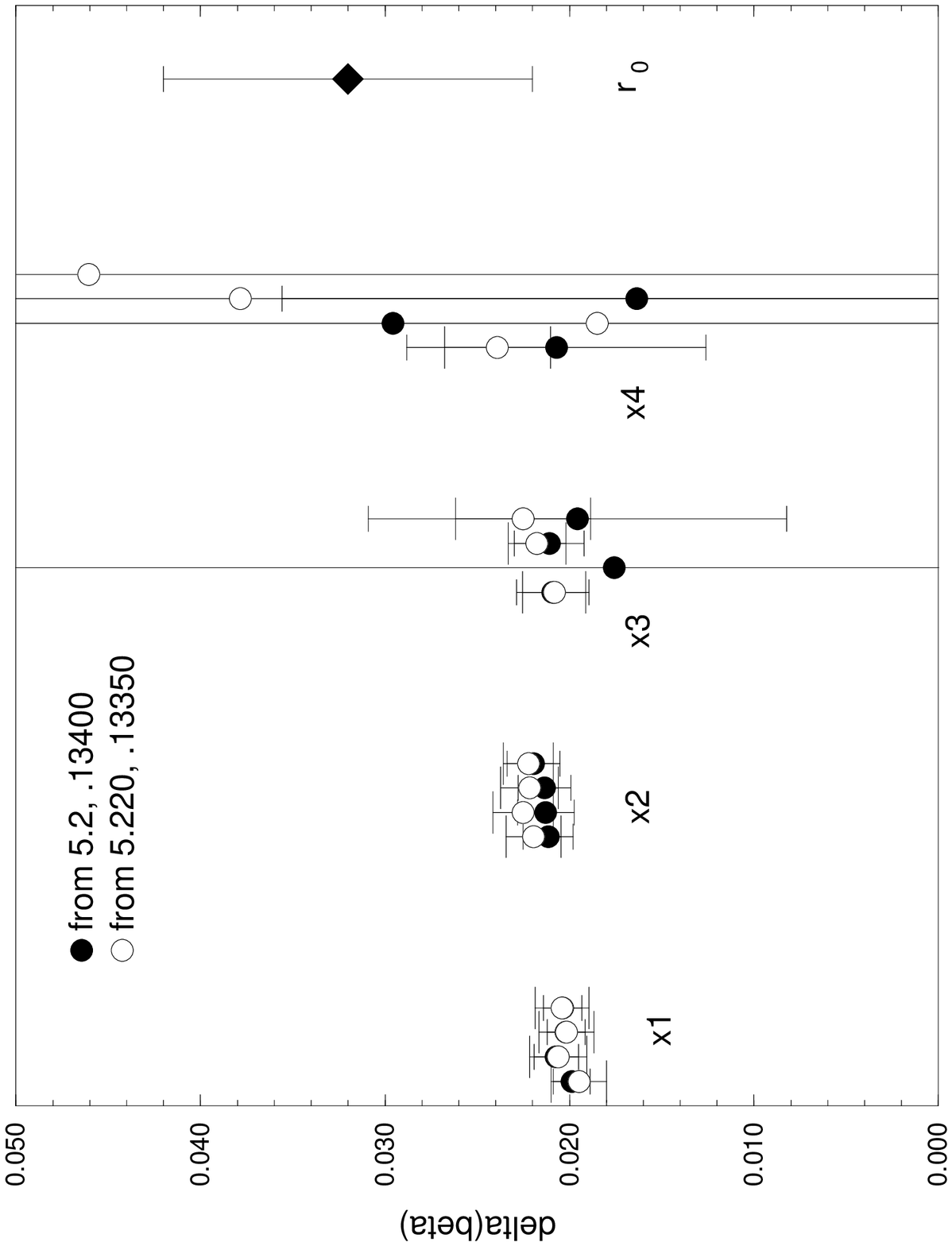}}
\vspace*{8pt}
\caption{Predictions for the shift $\delta\beta$ from 
reference point $(5.2,0.1340)$ 
obtained using the 16 sample loops described in the text.
The diamond point is the corresponding value deduced
from $r_0$. The closed (open) points correspond to a shift of
$\delta\kappa=-0.0005(+0.0005)$ from ensembles at 
$(5.2,0.1340)$ and  $(5.220,0.1335)$ respectively.
}
\label{fg:loopdbet}
\end{figure}

For each loop $L$, we have evaluated the shift $\delta\beta_L$
required to hold the Wilson loop $<W_L>$ constant under a change
$\delta\kappa$ ($=-0.0005$).
The results of this are shown 
in Figure~\ref{fg:loopdbet}.
The values of $\delta\beta_L$ 
are similar to each other and to that for the  average plaquette 
measured above. There is not much evidence of a shift increasing
with the loop size, although the $\times 2$ loops 
show a higher trend than the plaquette value.
One sees little evidence of mis-matching above the
level of a standard deviation.

If this result (same $\delta\beta$) was reproduced for {\em all}
gauge invariant loops and {\em all} linear combinations thereof, we
would conclude that the static potential, $r_0$ and hence the
lattice spacing, would be identical at the matched 
points~(\ref{eq:b0k0}) and~(\ref{eq:bk}). 
This would realise our initial 
objective of defining curves of constant effective volume.
However, it cannot be that {\em all} fixed-$F$ curves emanating from
a finite reference point $(\beta_0,\kappa_0)$ coincide.
Moving from this point, in one direction we approach the quenched limit
and in the other the chiral limit. We expect that different observables
will be more or less sensitive to the effects of quenching.
For example, the mass of a vector meson will probably change by 
less than 10\%{} as the chiral limit is reached in the full theory
as compared to its quenched value. On the other hand, the
string tension should change from the lattice equivalent of 440 Mev
to zero, eventually. We study the static potential in the next section.

\medskip
\subsection{Potential and $r_0$}
We have used the methods of~\cite{CMpot} to measure the potential
on each of the main ensembles studied. Since these are on $8^3\times 24$
lattices, there are strong finite size effects 
present in $V(r)$ and $r_0$ at the parameter values of
interest. However, for the present purpose
this is of little consequence. In the variational methods of ~\cite{CMpot}
one constructs \lq fuzzed\rq{} loops from a variety of spatial paths and 
employs transfer matrix
methods to extract energy eigenvalues.
The potential values were estimated by taking weighted
averages of the effective masses at large time. We took care to
use the same procedures on all ensembles. 
Errors were estimated by bootstrap.

\begin{figure}
\centerline{\includegraphics[width=\columnwidth]{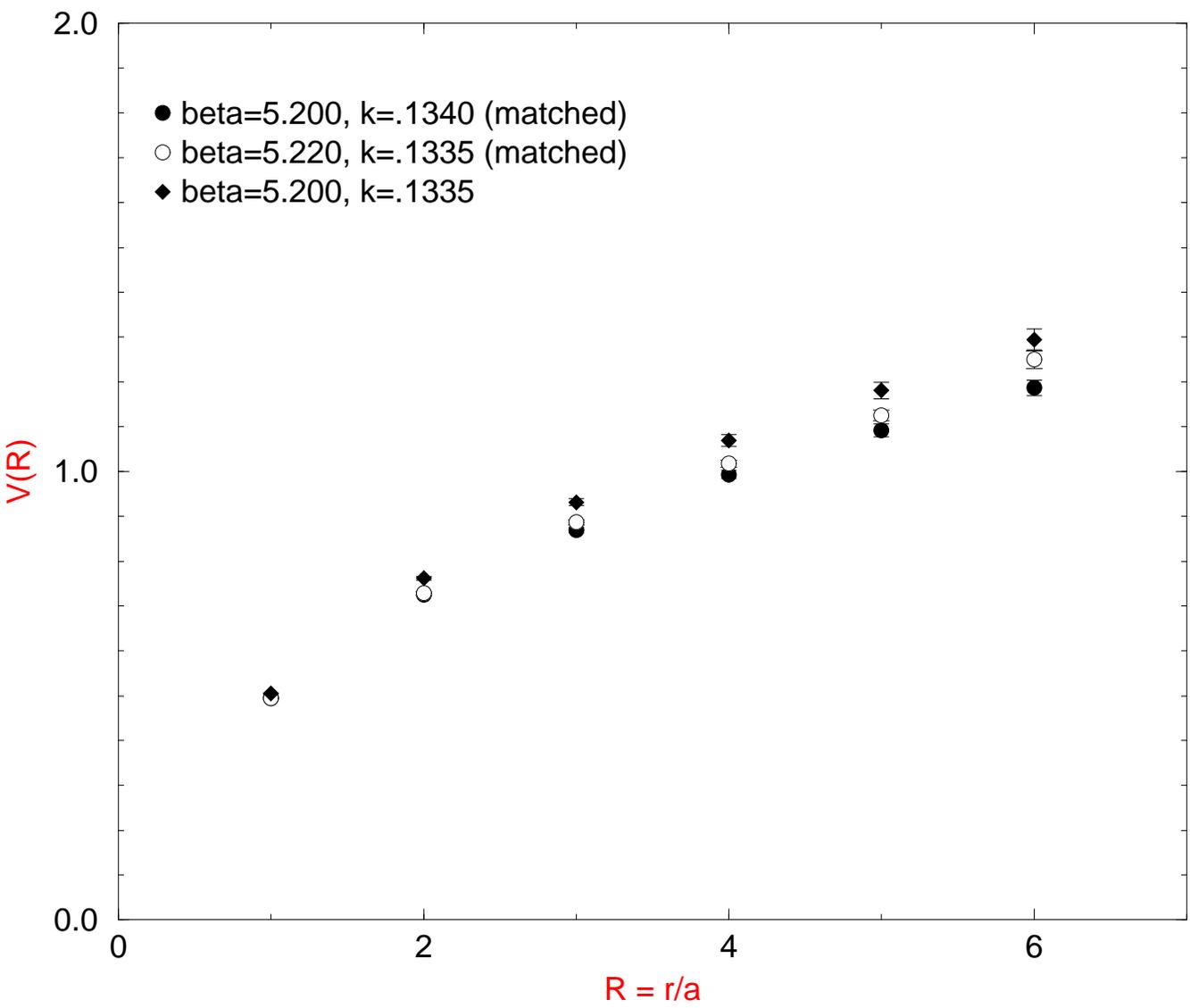}}
\vspace*{8pt}
\caption{Static potential on an $8^3\times 24$ lattice
at $(5.2,0.1340)$ and $(5.220,0.1335)$ where
the average plaquette values match.} 
\label{fg:potmatch}
\end{figure}

\begin{figure}
\centerline{\includegraphics[width=\columnwidth]{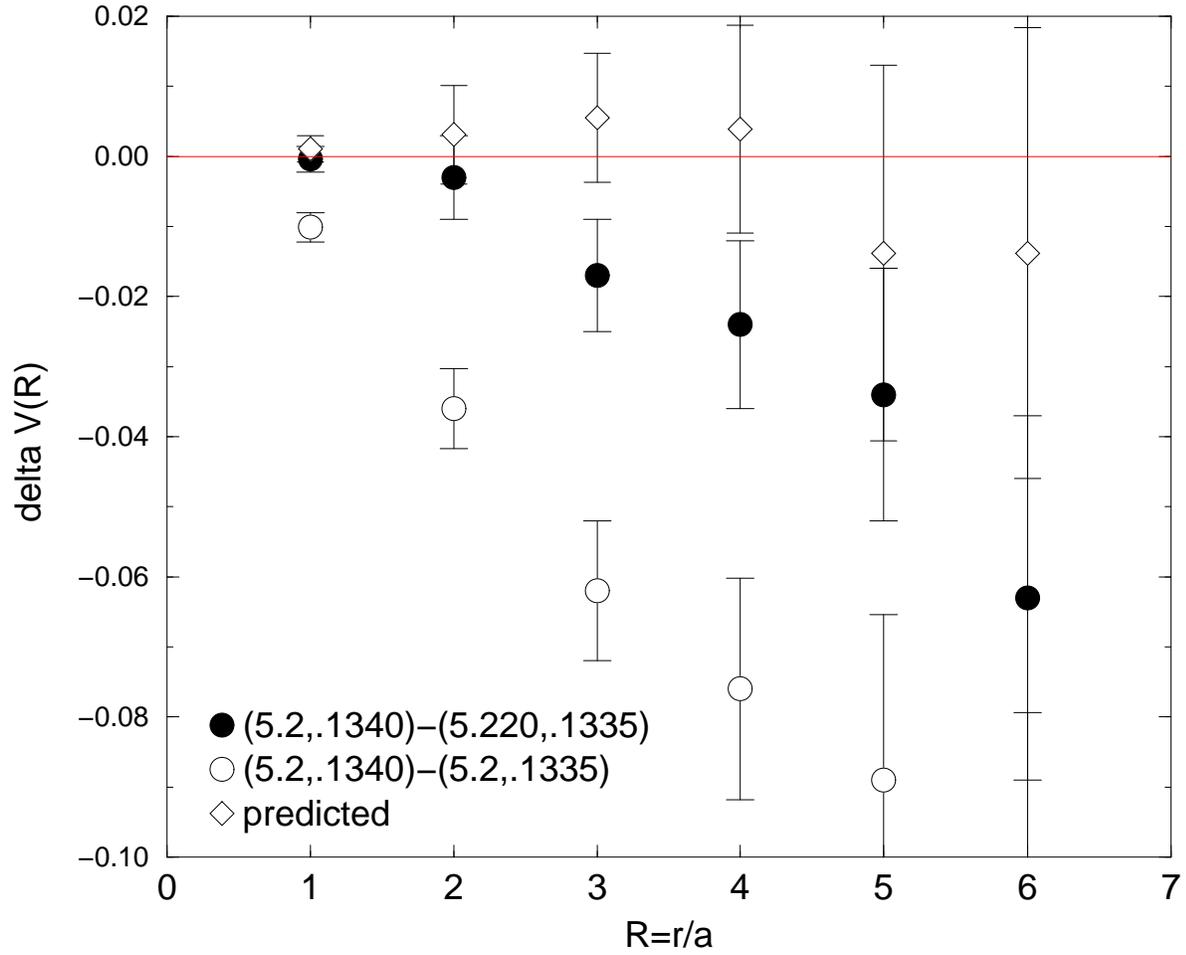}}
\vspace*{8pt}
\caption{Differences in the static potential measured from
the reference point $(5.2,0.1340)$. Solid and open circles correspond
to $(5.220,0.1335)$ and  $(5.2,0.1335)$ respectiveley.
Open diamonds corespond to the 
prediction for 
$(5.220,0.1335)$ using~(\ref{eq:F2F1}).
}
\label{fg:potdiff}
\end{figure}

Figure~\ref{fg:potmatch}  shows the static potential
at the matched points~(\ref{eq:b0k0}) and~(\ref{eq:bk}).
The values are in good agreement at short distances but show
a systematic divergence at larger separations.
Figure~\ref{fg:potdiff} shows more clearly 
the difference between the potentials $V(5.2,0.1340)-V(5.220,0.1335)$.
There appears to be a systematically increasing difference at larger
distance. For comparison, the figure also shows 
$V(5.2,0.1340)-V(5.2,0.1335)$ where there has been a shift in $\kappa$
but {\em no} compensating change in $\beta$. There is clear disagreement
at all distances, as expected. 

The remaining set of points in
Figure~\ref{fg:potdiff} shows the prediction for
$(5.220,0.1335)$ from the reference ensemble at $(5.2,0.13400)$
using~(\ref{eq:F2F1}) to first order. 
Within the large statistical
errors, the predicted difference is indeed compatible with zero.
This demonstrates that where matching has been done only approximately 
(in this case with $<P>$) an observable can still be 
reliably estimated at another nearby point of interest.  

From the comparison of the directly simulated points, 
we conclude that matching the plaquette is not equivalent to
matching the long range potential. Note that no corrections have been made for
lattice artifacts at short distances but these are expected to be similar
in each case.

From the potential measurements represented in
Figure~\ref{fg:potmatch} we can extract corresponding
values of $r_0$~\cite{R0} and, hence,
lattice spacing using $r_0({\rm phys})=0.49\, {\rm fm}$.  
These are shown in Table~\ref{tb:r0}.
In finding $r_0$, we have calculated interpolated values of
the static force via Newtonian 5-point interpolation of
the potential. As usual, errors were computed via bootstrap.
The same procedures were used for each ensemble.

From Table~\ref{tb:r0}, we see that the lattice volume
at the matched points is similar although that at the heavier
quark mass is perhaps one standard deviation larger.
This is a reflection of the observation made above that 
the potentials diverge at large separations due to their different
slopes.
We have attempted to estimate $\delta\beta$ required to match the
mean values of the lattice spacings. In the last three rows
of Table~\ref{tb:r0}, we show the lattice spacing predicted
from the reference ensemble via (\ref{eq:F2F1}) for $\beta$ shifts
of $+0.02$, $+0.03$ and $+0.04$ to go with the $\kappa$ shift of
$-0.0005$. The spacing predicted for $\delta\beta=0.02$ agrees well
with that measured by direct simulation (previous line).
From this value and those corresponding to $\delta\beta=0.03$ and $0.04$ we
estimate that the optimal matching for the
lattice spacing (as opposed to the plaquette) would be
$\delta\beta=0.032(10)$. This value is compared with those corresponding
to the various Wilson loops in Figure~\ref{fg:loopdbet}.

We conclude that the constant lattice volume curve (as defined by
$r_0$) may indeed differ from that corresponding to constant plaquette
value. With only 40 configurations,
the evidence is of marginal statistical significance.
As a cross-check we measured $r_0$ directly from a simulation
at $(5.232,.1335)$. See Table~\ref{tb:r0}. The result was compatible 
with the prediction and with the matched ensemble from which the
prediction was made. Again, the statistical significance is not high.

We have also measured $r_0$ on quenched configurations at
$\beta=5.61$, the point which is
predicted and verified to have matched values
of $<P>$, and at $\beta=5.85$, close to the point
where $r_0$ is expected to match. Results are shown in Table~\ref{tb:avP}.
As expected, the lattice spacing does {\em not} match
at $5.61$ but is close to matching at $5.85$. The shift
required to match $r_0$ is some $60\%$ larger than that required to
match $<P>$.

\subsection{Hadron correlators}
\begin{figure}
\centerline{\includegraphics[width=\columnwidth]{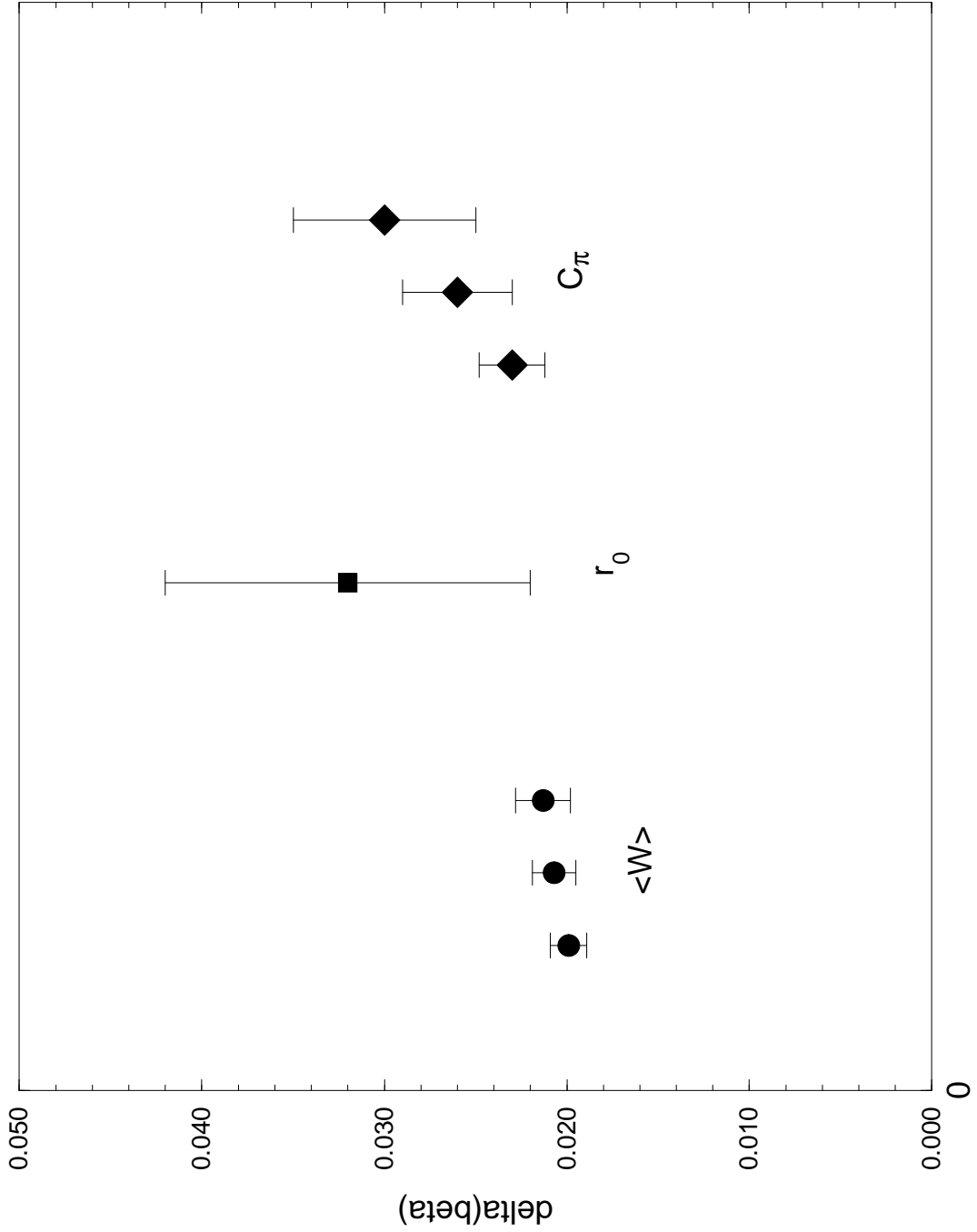}}
\vspace*{8pt}
\caption{Predictions for the shift $\delta\beta$ from 
reference point $(5.2,0.1340)$ required to keep
the specified observables constant when $\kappa$
is changed by $\delta\kappa=-0.0005$.
Circles correspond to Wilson loops ($1\times 1$,
$1\times 2$, $2\times 4$); 
diamonds to $C_{\pi}(t)$ ($t=0$, $1$, $2$);
square to $r_0$.}
\label{fg:gendbet}
\end{figure}

Finally in this section, we present results from matching
lattice pion correlators. This test is made more practicable by recent
advances in measuring 
hadron correlators with good statistical precision on a single
gauge configuration~\cite{CMalltoall}.
We have made measurements of the local pion correlator
$C_{\pi}(t)$ on the reference ensemble and calculated the corresponding
$\delta\beta$ shifts. That is, for each value of $t$, we estimate
the shift $\delta\beta$ required to keep $C_{\pi}$ constant for 
the test $\kappa$ shift. 
Results are presented Table~\ref{tb:db} where they 
are compared with examples of other observables.
At short time separations ($0$ and $1$)
we find values compatible with that for the plaquette. At large
values of $t$ the results are overwhelmed by noise and we
are unable to draw conclusions. However there are indications
that for increasing $t$ the shift required in $\beta$ is also increasing.
See, for example the correlator for $t=2$ and corresponding effective mass
values which are also shown in the table:
\beq
m_{\pi}^{\rm eff}(t)=-\ln[C_{\pi}(t+1)/C_{\pi}(t)]\, .
\eeq
Clearly one requires greater statistics to confirm these trends, but
they are consistent with those inferred 
from the $r_0$ results presented above. 
A compilation of shifts $\delta\beta$ for sample loops, $C_{\pi}(t)$  and
$r_0$ are shown in Figure~\ref{fg:gendbet}.

Note that the effective mass
values involved in this test are very far from those of a physical pion.

\section{Further applications}
\label{s:applics}
The above matching technology has a variety of potential uses
including the following.

\subsection{Parallel tempering}
\label{s:pt}
Parallel tempering (PT) is an improved Monte--Carlo method originally 
proposed by {\em Hukushima} et al~\cite{Hukushima} 
to improve simulations of spin glasses.
It was further discussed by {\em Marinari et al} in \cite{Marinari} and
\cite{MariPari2} who suggested its use in Lattice QCD.  Recently 
{\em Boyd}~\cite{Boyd} 
applied the technique to lattice QCD with staggered
fermions and found evidence that Parallel Tempering did help decorrelate 
long distance observables. 

Parallel tempering essentially consists of running several independent 
simulations in parallel and with different parameters. Each such 
simulation produces a set of configurations which is distributed according
to the probability distribution dictated by the simulation action and 
parameters. The PT algorithm exploits the fact that these distributions
may have an overlap and occasionally attempts to swap configurations 
between ensembles. Acceptance of the swap is controlled by a Metropolis
style acceptance step. 

This is the same situation described by the matching criterion 
M3 in section~\ref{s:curves}. From another viewpoint, 
the distance between the actions in matching
criterion M2 can be related to the acceptance rate of the swaps in a 
parallel tempering algorithm.

In \cite{Boyd}, tempering was carried out
in the quark mass only. All the ensembles had the same value of $\beta$.
Furthermore, the quark masses had to be spaced quite closely together to 
obtain a reasonable swap acceptance rate. With 
the matching technology presented in this paper, 
it is possible to temper in {\em both} $\beta$ and 
the quark mass.
This may allow one to perform PT along a curve
of approximately constant volume, 
and at a such a separation between ensembles that 
one might use some of the tempering ensembles to perform 
chiral extrapolations. Alternatively, one might be able to simulate
with ensembles suitably chosen to improve the decorrelation properties
of the system as a whole.
A detailed investigation into PT using
the matching technology is being conducted by us and the full
results will be reported elsewhere \cite{PTpaper}.

\subsection{Approximate algorithms}
In~\cite{ACIJCS} we demonstrated how the parameters of approximate
algorithms could be tuned according to the criteria M1, M2 or M3
described above. In subsequent tests~\cite{ACIlat97}, we 
showed that approximate algorithms based on a few Wilson loops only,
were unlikely to produce a very accurate approximation, at least in the
sense of M3 where the approximate action acts as the guide
within an exact algorithm. The variance of the difference between the two 
extensive quantities remains unacceptably large on lattices of a useful
size. It is, however, still of considerable interest to design
approximate or model actions which improve on the quenched approximation
by encapsulating at least some of the additional physics implied by
dynamical quarks. 

An alternative route to an exact algorithm might be to make use of the
Lanczos quadrature approximation of section~\ref{s:trln}  for part of
the effective action and gauge invariant loops as above for the
remainder. The trial configurations would be generated by the loop part
of the action and an accept/reject step based on the Lanczos part. The
technology of section~\ref{s:curves} can be used to tune the loop part
together with the Lanczos part to match the exact action. 
As a simple example of this approach, consider an approximate action
defined such that (c.f. (\ref{eq:Seff}))
\beq
S_{\rm app}=-\beta' W_{\Box}-T(N_{\rm Lanc})
\label{eq:Sapp}
\eeq
where $T(N_{\rm Lanc})$ is the approximation to
$\Tr\ln (M^{\dagger}M)$ as described in section~\ref{s:trln} but using
only $N_{\rm Lanc}$ Lanczos iterations. 
The loop part of the effective action is just a single
plaquette in this example. In a standard Metropolis
update this action would only be viable if $N_{\rm Lanc}$ was considerably smaller
than the typical values (90) required to estimate the true action, and
sufficient account was taken of the short range fluctuations by
having a properly tuned gauge loop part. 
This approximation is similar in spirit to that advocated
in~\cite{Bardeen,Duncan} where it is argued that a truncated sum
of low-lying eigenvalues can
reproduce the gross behaviour of the fermion determinant.
In the present scheme, we are able to obtain a particularly efficient
approximation to the trace log by using the optimal weighting
determined by the Gaussian quadrature rule. We have conducted
preliminary tests of these ideas by measuring the shift 
$\delta\beta=\beta'-\beta$
required to compensate for a truncation to $N_{\rm Lanc}$
Lanczos iterations. In a simple test which matched the average 
plaquette, the shift in $\beta$ was reduced by a factor of $8$
in changing from $N_{\rm Lanc}=0$ (quenched) to $N_{\rm Lanc}=4$, for
example. The correspond residual variance (after matching) also dropped 
by a factor of around $8$. For increasing
$N_{\rm Lanc}$, the shift rapidly becomes compatible with
zero at the level of statistical accuracy implied by the number of
noise vectors used ($80$). 
This demonstrates that the Lanczos quadrature approach
gives a very efficient estimator for the trace log.
It will be worth exploring whether 
the long range modes described by such an approximation can be
combined with a suitably tuned gauge loop action describing
short range modes so as
to achieve a practical exact algorithm.

\section{Conclusions}
We have proposed a strategy for dynamical quark simulations in which
the effective lattice volume is held fixed while the effects of
progressively lighter sea quarks are investigated. Possible algorithms
for accomplishing this have been presented and the results of tests
discussed. In particular, we have presented results using an efficient
stochastic estimator of the fermion determinant and quantities related
to it. These include estimates of the constant lattice spacing curves
at relevant points in the $\beta,\kappa$ plane for lattice QCD using
a non-perturbatively improved action. We have demonstrated that the work
involved in determining such curves via our differential stochastic
methods is considerably less than that required to establish them by
direct simulation. Further applications of these
techniquess have been discussed.

\acknowledgements
{Computational resources for this work were in part provided by 
the HPCI inititaive of EPSRC under grant GR/K41663.
Alan Irving and James Sexton are grateful to the British Council/
Forbairt Joint Research Scheme for travel support.
James Sexton would also like to thank Hitachi Dublin Laboratory for its support.
}



 
\newpage


\begin{table*}
\caption{Parameters association with HMC simulation and with the
stochastic estimation of the determinant.}
\begin{tabular}{lll}
Parameter &Description &Value	\\
\hline
$N_{\rm step}$ &number of HMC steps per trajectory &50\\
$N_{\rm solve}$ &number of sweeps in the HMC solver (e.g.BiCGStab) &300\\
$A$ &HMC trajectory acceptance &0.75\\
$\tau_{\rm AC}$ &autocorrelation time in trajectories &30\\
\hline
$N_{\phi}$ &number of noise vectors (section~\ref{s:trln}) &80\\
$N_{\rm Lanc}$ &number of Lanczos iterations &90\\
\end{tabular}
\label{tb:work}
\end{table*}

\begin{table*}
\caption{Sample variances usued to estimate the relative work
involved in measuring differences directly and by the techniques
proposed in this paper. These correspond to a shift
$\delta\kappa=-.0005$ and the reference data described in the text.}
\begin{tabular}{llll}
$F$ &$\sigma_F^2$ &$\sigma^2_{\delta F}$ &$\sigma^2_{\delta F}/\sigma_F^2$\\
\hline
$P$ (average plaquette)	&$5.2\times 10^{-6}$	&$2.9\times 10^{-5}$ &$5.5$\\
$r_0$ 		&$0.78$	&$2.7$ &$3.5$\\
\end{tabular}
\label{tb:vars}
\end{table*}


\begin{table*}
\caption{Matching the average plaquette.}
\begin{tabular}{lrrrllr}
$(\beta_0,c_{\rm sw},\kappa_0)$ & $<P>$  & traj   
	& $\kappa$  &$\phantom{xx}\delta\kappa$   &$\phantom{xx}\delta\beta$ & config\\ 
\hline
$(5.200,2.0171,0.1340)$ & $0.5286(3)$   &$6000$ 
	& $0.1335$    &$-0.0005$    & $+0.0199(10)$ &  $40$\\
	&&& $0$	&& $+0.41(3)$	&  $40$\\
$(5.220,1.9936,0.1335)$	& $0.5290(3)$   & $6000$ 
	& $0.1340$  & $ +0.0005$ & $-0.0195(15)$ &   $40$\\
	&&& $0.1330$  & $-0.0005$  &   $ +0.0162(23)$ &   $40$\\
	&&& $0$ && $+0.40(4)$	&  $40$\\
$(5.61,0)$	& $0.5275(3)$   &$1000$ sweeps\\ 
\hline
$(5.200,2.0171,0.1330)$	& $0.5197(3)$   & $6000$
	& $0.1325$    	& $+0.015(3)$  && $40$\\
	&&&  $0$	& $+0.368(16)$	&&  $40$\\
$(5.215,1.9994,0.1325)$	& $0.5207(6)$   & $2000$ &&&\\
\end{tabular}
\label{tb:avP}
\end{table*}


\begin{table}
\caption{Construction of sample gauge invariant loops
from specified link steps.
A step $+2$ means a link along lattice direction $2$ while
$-1$ means a link along lattice axis $1$ in the negative direction.}
\label{tb:loops}
\begin{tabular}{ccl}
Loop & No. of links & Link steps\\
\hline
1   & 4 &(+1,+2,-1,-2) \\
2   & 6 &(+1,+1,+2,-1,-1,-2)\\
3   & 6 &(+1,+2,+3,-2,-1,-3)\\
4   & 6 &(+1,+2,+3,-1,-2,-3)\\
\end{tabular}
\end{table}


\begin{table}
\caption{Values of $r_0$ and lattice spacing
deduced from the static potential. The last two
rows are values predicted by (\ref{eq:F2F1}) from
the reference ensemble $(5.2,0.13400)$.
}
\label{tb:r0}
\begin{tabular}{llll}
$(\beta,\kappa)$ & $r_0/a$ &a {\rm fm} &\phantom{xxxxxxxx}\\
\hline
$(5.200,0.1340)$  &$3.87(17)$   &$0.127(6)$   &direct simulation\\
$(5.220,0.1335)$  &$3.69(11)$   &$0.133(4)$   &\phantom{x}by HMC\\
$(5.232,0.1335)$  &$3.76(13)$   &$0.130(4)$   &\\
$(5.61,0)$  	&$2.33(2)$   &$0.211(2)$   &\\
\hline   
$(5.220,0.1335)$  &$3.67(18)$   &$0.134(6)$   &pred. \\
$(5.230,0.1335)$  &$3.80(22)$   &$0.129(9)$   &from $(5.2,0.13400)$\\
$(5.240,0.1335)$  &$4.06(33)$   &$0.121(12)$  &\phantom{x}using~(\ref{eq:F2F1})\\   
\end{tabular}
\end{table}


\begin{table}
\caption{Summary of the shifts $\delta\beta$ required to 
maintain the specified quantities constant under a change
$\delta\kappa=-0.0005$ from the reference point
$(\beta_0,\kappa_0)=(5.200,0.13400)$. 
}
\label{tb:db}
\begin{tabular}{lll}
Measurement 	& Value at $(\beta_0,\kappa_0)$ &$\delta\beta$\\
\hline
$<P>$			&$0.5286(3)$	&$0.0199(10)$\\
$<W_{1\times 2}>$	&$0.3161(4)$	&$0.0207(12)$\\
$<W_{2\times 4}>$	&$0.0312(2)$	&$0.0213(15)$\\
$r_0/a$			&$3.87(17)$	&$0.0302(10)$\\
$C_{\pi}(0)$		&$2.132(2)$	&$0.0230(18)$\\
$C_{\pi}(1)$		&$0.2782(8)$	&$0.026(3)$\\
$C_{\pi}(2)$		&$0.0792(5)$	&$0.030(5)$\\
$C_{\pi}(3)$		&$0.0298(3)$	&$0.035(80)$\\
$m_{\rm eff}^{\pi}(1)$	&$2.037(2)$	&$0.030(5)$\\
$m_{\rm eff}^{\pi}(2)$	&$1.256(4)$	&$0.035(28)$\\
\end{tabular}
\end{table}

\end{document}